\documentstyle[epsfig,12pt,a4p]{article}

\newcommand{\piz}{\mbox{$\pi^0\pi^0\pi^0\pi^0$ }}
\newcommand{\pic}{\mbox{$\pi^+\pi^-\pi^+\pi^-$ }}
\newcommand{\pim}{\mbox{$\pi^+\pi^-\pi^0\pi^0$ }}
\newcommand{\etapipi}{\mbox{$\eta \pi^+\pi^-$ }}

\begin{document}
\begin{titlepage}
\def\footnoterule{\hrule width 1.0\columnwidth}
%\hfill  \hfill
%6\thinspace February\thinspace 1990
% \begin{center} {\large EUROPEAN ORGANIZATION FOR NUCLEAR RESEARCH}
%  \end{center}
\begin{tabbing}
put this on the right hand corner using tabbing so it looks
 and neat and in \= \kill
\> {2 December 1999}
\end{tabbing}
\bigskip
\bigskip
\begin{center}{\Large  {\bf A spin analysis of the
$4\pi$ channels produced in central
pp interactions at 450 GeV/c}
}\end{center}

\bigskip
\bigskip
\begin{center}{        The WA102 Collaboration
}\end{center}\bigskip
\begin{center}{
D.\thinspace Barberis$^{  4}$,
F.G.\thinspace Binon$^{   6}$,
F.E.\thinspace Close$^{  3,4}$,
K.M.\thinspace Danielsen$^{ 11}$,
S.V.\thinspace Donskov$^{  5}$,
B.C.\thinspace Earl$^{  3}$,
D.\thinspace Evans$^{  3}$,
B.R.\thinspace French$^{  4}$,
T.\thinspace Hino$^{ 12}$,
S.\thinspace Inaba$^{   8}$,
A.\thinspace Jacholkowski$^{   4}$,
T.\thinspace Jacobsen$^{  11}$,
G.V.\thinspace Khaustov$^{  5}$,
J.B.\thinspace Kinson$^{   3}$,
A.\thinspace Kirk$^{   3}$,
A.A.\thinspace Kondashov$^{  5}$,
A.A.\thinspace Lednev$^{  5}$,
V.\thinspace Lenti$^{  4}$,
I.\thinspace Minashvili$^{   7}$,
J.P.\thinspace Peigneux$^{  1}$,
V.\thinspace Romanovsky$^{   7}$,
N.\thinspace Russakovich$^{   7}$,
A.\thinspace Semenov$^{   7}$,
P.M.\thinspace Shagin$^{  5}$,
H.\thinspace Shimizu$^{ 10}$,
A.V.\thinspace Singovsky$^{ 1,5}$,
A.\thinspace Sobol$^{   5}$,
M.\thinspace Stassinaki$^{   2}$,
J.P.\thinspace Stroot$^{  6}$,
K.\thinspace Takamatsu$^{ 9}$,
T.\thinspace Tsuru$^{   8}$,
O.\thinspace Villalobos Baillie$^{   3}$,
M.F.\thinspace Votruba$^{   3}$,
Y.\thinspace Yasu$^{   8}$.
%% \end authorlist
}\end{center}

\begin{center}{\bf {{\bf Abstract}}}\end{center}

{
The reactions
$ pp \rightarrow p_{f} (X^0) p_{s}$, where $X^0$ is observed
decaying to  \piz, \pic and \pim,
have been studied at 450 GeV/c.
There is evidence for an $a_2(1320) \pi$ decay mode of the
$\eta_2(1645)$ and $\eta_2(1870)$ in the \pim and \pic
final states.
The $f_2(1950)$ is consistent with being a single resonance with
a dominant $f_2(1270) \pi \pi $ decay mode.
The $f_0(1370)$ is found to decay dominantly to $\rho \rho$ while
the $f_0(1500)$ is found to decay to $\rho \rho$ and $\sigma \sigma$.
}
\bigskip
\bigskip
\bigskip
\bigskip\begin{center}{{Submitted to Physics Letters}}
\end{center}
%\newpage
\bigskip
\bigskip
\begin{tabbing}
aba \=   \kill
% $^\dag$ \> \small
% Deceased. \\
$^1$ \> \small
LAPP-IN2P3, Annecy, France. \\
$^2$ \> \small
Athens University, Physics Department, Athens, Greece. \\
%% $^3$ \> \small
%% Bergen University, Bergen, Norway. \\
$^3$ \> \small
School of Physics and Astronomy, University of Birmingham, Birmingham, U.K. \\
$^4$ \> \small
CERN - European Organization for Nuclear Research, Geneva, Switzerland. \\
$^5$ \> \small
IHEP, Protvino, Russia. \\
$^6$ \> \small
IISN, Belgium. \\
$^7$ \> \small
JINR, Dubna, Russia. \\
$^8$ \> \small
High Energy Accelerator Research Organization (KEK), Tsukuba, Ibaraki 305-0801,
Japan. \\
$^{9}$ \> \small
Faculty of Engineering, Miyazaki University, Miyazaki 889-2192, Japan. \\
$^{10}$ \> \small
RCNP, Osaka University, Ibaraki, Osaka 567-0047, Japan. \\
$^{11}$ \> \small
Oslo University, Oslo, Norway. \\
$^{12}$ \> \small
Faculty of Science, Tohoku University, Aoba-ku, Sendai 980-8577, Japan. \\
\end{tabbing}
\end{titlepage}
\setcounter{page}{2}
\bigskip
\par
The WA76, WA91 and WA102 collaborations
have studied the centrally produced
$\pi^{+}\pi^{-}\pi^{+}\pi^{-}$ final state in the reaction
\begin{equation}
pp \rightarrow p_{f} (\pi^{+}\pi^{-}\pi^{+}\pi^{-}) p_{s}
\label{eq:a}
\end{equation}
at 85~\cite{re:f},
300~\cite{re:a},
and 450 GeV/c \cite{re:wa914pi,re:wa1024pi}.
The subscripts $f$ and $s$ indicate the
fastest and slowest particles in the laboratory respectively.
In addition to the $f_1(1285)$, which was observed at all energies, peaks
were observed
at 1.45 and 1.9 GeV in the 300 and 450~GeV/c data.
In contrast,
no clear evidence was seen for these states in the 85~GeV/c
data of the same experiment~\cite{re:f}.
The increased prominence of these states with increased
incident energy~\cite{re:a}
is consistent with the formation of these states via a
double Pomeron exchange mechanism, which is predicted
to be a source of
gluonic states \cite{re:b}.
\par
The peak at 1.9 GeV,
called the $f_{2}(1950)$, was found to have
$I^{G}J^{PC}=0^{+}2^{++}$ and decay to
$f_2(1270)\pi \pi$ and $a_2(1320)\pi$~\cite{re:wa914pi,re:wa1024pi}.
It was not
possible to determine whether this was one resonance
with two decay modes, or two resonances.
However, in a recent analysis of the centrally
produced \etapipi final state by the WA102 experiment~\cite{etapipi},
no evidence was found for a $J^{PC}$~=~$2^{++}$ $a_2(1320)\pi$ decay mode.
The peak at 1.45 GeV was shown to have
$I^{G}J^{PC}=0^{+}0^{++}$ and decay to $\rho \rho $.
It has been described
as being due to the interference between the
$f_0(1370)$ and the $f_0(1500)$
 \cite{re:wa914pi,re:wa1024pi} which implies that
both states should have a substantial $\rho \rho$ decay mode.
However,
the initial analysis of the $4\pi$ channel in $p \overline p$ by
Crystal Barrel~\cite{cb4pi}
showed that the $f_0(1500)$ decays via
$\sigma \sigma$ with a very weak coupling to $\rho \rho$.
A recent preliminary analysis~\cite{thoma} of the
Crystal Barrel data concludes that the $f_0(1370)$ decays
strongly to $\sigma \sigma$ and $\rho \rho$ while the
$f_0(1500)$ decays to $\sigma \sigma$ and $\pi^* \pi$ where
the $\pi^* \pi$ decays to $\rho \pi$.
\par
Therefore, there are still several unanswered questions on the
decays of the $f_2(1950)$, $f_0(1370)$ and $f_0(1500)$.
In order to address these questions, in this paper
an analysis of several different $4\pi$ decay modes will be presented.
In particular, to investigate the $\sigma \sigma$ contribution
in the decays of the $f_0(1370)$ and $f_0(1500)$.
One of the major problems of studying the \pic final state
is the number of possible isobar decay modes that
are present. Therefore
a study will be made firstly
of the \piz final state which
has the nice feature that it is
free from any contribution from
$\rho(770)$, $a_1(1260)$ or $a_2(1320)$ isobars.
Next
an analysis will be presented of the
\pim decay mode which has different systematic effects
compared to the \pic decay.
Finally a reanalysis of the \pic decay mode will be presented.
\par
The data come from the WA102 experiment
which has been performed using the CERN Omega Spectrometer,
the layout of which is
described in ref.~\cite{WADPT}.
The reaction
\begin{equation}
pp \rightarrow p_{f} (\pi^0 \pi^0\pi^0 \pi^0) p_{s}
\label{eq:e}
\end{equation}
has been studied at 450~GeV/c.
Reaction~(\ref{eq:e})
has been isolated
from the sample of events having two
outgoing
charged tracks and eight $\gamma$s reconstructed in the GAMS-4000
calorimeter,
by first imposing the following cuts on the components of
missing momentum:
$|$missing~$P_{x}| <  14.0$ GeV/c,
$|$missing~$P_{y}| <  0.20$ GeV/c and
$|$missing~$P_{z}| <  0.16$ GeV/c,
where the $x$ axis is along the beam
direction.
A correlation between
pulse-height and momentum
obtained from a system of
scintillation counters was used to ensure that the slow
particle was a proton.
\par
Fig.~\ref{fi:1}a) shows the two photon mass spectrum
for $8\gamma$-events  when
the mass of the other three $2\gamma$-pairs lies
within a band around the $\pi^0$ mass (100--170 MeV).
A clear $\pi^0$ signal is observed with a small background.
Events belonging to
reaction (\ref{eq:e}) have been selected using a
kinematical fit (8C fit, four-momentum
conservation being used and the masses of four $\pi^0$s being fixed).
Events corresponding to the reaction $\eta \pi^0$, where the
$\eta$ decays to $3\pi^0$, have been removed by requiring
$M(3\pi^0)$~$\ge$~0.6~GeV.
The major background to the \piz final state comes from the
decay of the $\eta^\prime$ and $f_1(1285)$ to $\eta \pi^0 \pi^0$
where the $\eta$ decays to $3\pi^0$ and one $\pi^0$ is undetected.
This produces events in the threshold to 1.3~GeV region of the
\piz mass spectrum.
A kinematical fit has been used to remove these events.
Using real $\eta \pi \pi$ events a simulation has been performed
to estimate the residual contamination in the \piz final state
which is found to be less than 5~\%.
\par
The resulting \piz effective mass spectrum is shown in
fig.~\ref{fi:1}b) and consists of 1438 events. There is a broad enhancement
in the 2~GeV region, compatible with the $f_2(1950)$,
but there is no clear peak in the
1.45~GeV region.
\par
One of the aims of this paper is to investigate the role of the $\sigma$
in the decays of the $4\pi$ final state.
In order to determine the amount of $\sigma \sigma$ a parameterisation of the
$\pi\pi$ S-wave ($\sigma$) is required.
Parameters for the $\sigma$ may
be process dependent.
The nature of the $\sigma$ is not clearly established
and parameterisations of it may be process dependent.
It may be a resonant state or just two pions produced in relative S-wave.
However, the parameterisation of the $\pi \pi$ S-wave
is crucial in all
these analyses in order to determine the amount of $\sigma \sigma$ observed.
Assuming that the 1.3$-$1.5 GeV \piz mass region is dominated by
the decay to $\sigma \sigma$, then
in order to determine the best parameters for the $\sigma$,
a study has been made of the $\pi^0\pi^0$ mass spectrum
for the region 1.3~$\leq$~M(\piz)~$\leq$~1.5~GeV, shown in
fig.~\ref{fi:1}c).
A Monte Carlo simulation has been performed using the decay
$X^0$~$\rightarrow$~$\sigma \sigma$ with $\sigma$~$\rightarrow$
$\pi^0\pi^0$.
Formally, we describe the shape of the $\sigma$
using a spin 0 relativistic
Breit-Wigner amplitude of the form:
\[
A(M_{\pi \pi }) = \left( \frac{q}{q_0}\right)
\frac{\Gamma m_0}{m_0^2 - m_{\pi \pi}^2 -im_0\Gamma}
\]
In order to describe the
centrally produced
mass spectrum the resulting function $|A(M_{\pi\pi})|^2$ has been multiplied by
the
kinematical factor $(M_{\pi\pi} - 4m_\pi^2)^{1/2}/M_{\pi\pi}^3$~\cite{re:AMP}.
%However, it should be pointed out that
%any other function of a similar shape would be
%equally good for the following analysis.
The mass and width of the $\sigma$ have been
varied and the Monte Carlo data passed through the detector simulation
and the resulting mass distributions were compared to the real data.
The parameters that best describe the data are
$m_0$~=~450~$\pm$~50~MeV and $\Gamma$~=~600~$\pm$~50~MeV
and the resulting mass spectrum (solid histogram)
is superimposed on the real $\pi^0\pi^0$
mass spectrum shown in fig.~\ref{fi:1}c).
% This shape is different from that found in the
% centrally produced $\pipi$ mass spectrum which peaks nearer
% to threshold~\cite{pipipap}.
\par
A spin-parity analysis of the
\piz
channel
has been performed in 120~MeV \piz mass bins
using an isobar model and the method described in ref.~\cite{re:wa914pi},
assuming that
only
the $\sigma \sigma$ and $f_2(1270) \pi \pi$ intermediate
states with $J^{PC}$~=~$0^{++}$ or $2^{++}$ contribute.
The data can be fitted using only the
$J^{PC}$~=~$0^{++}$ $\sigma \sigma$ wave and the
$J^{PC}$~=~$2^{++}$ $f_2(1270) \pi \pi$ wave with $J_Z$~=~0 which
are shown in figs.~\ref{fi:1}d) and e) respectively.
Superimposed on fig.~\ref{fi:1}d) is the distribution that
would be expected
for the $f_0(1500)$; as can be seen there is little
evidence for any $f_0(1370)$ contribution.
Superimposed on fig.~\ref{fi:1}e) is what would be expected
from the $f_2(1950)$; as can be seen it well describes
the
$J^{PC}$~=~$2^{++}$ $f_2(1270) \pi \pi$ wave, showing that
the peak at 1.95 GeV is consistent with being a single resonance.
\par
An analysis has next been performed on the centrally
produced \pim system.
The reaction
\begin{equation}
pp \rightarrow p_{f} (\pi^+ \pi^-\pi^0 \pi^0) p_{s}
\label{eq:g}
\end{equation}
has been isolated
from the sample of events having four
outgoing
charged tracks and four $\gamma$s reconstructed in the GAMS-4000
calorimeter,
by first imposing the following cuts on the components of the
missing momentum:
$|$missing~$P_{x}| <  17.0$ GeV/c,
$|$missing~$P_{y}| <  0.16$ GeV/c and
$|$missing~$P_{z}| <  0.12$ GeV/c.
The two photon mass spectrum (not shown)
when
the mass of the other $2\gamma$-pair lies
within a band around the $\pi^0$ mass (100--170 MeV) shows
a clear $\pi^0$ signal with small background.
Events belonging to
reaction (\ref{eq:g}) have been selected using a
kinematical fit (6C fit, four-momentum
conservation being used and the masses of two $\pi^0$s being fixed).
Events containing a fast $\Delta^{++}(1232) $
were removed if $M(p_{f} \pi^{+}) < 1.3 $ GeV, which left
283 408 centrally produced \pim events.
\par
Fig.~\ref{fi:3}a shows the \pim
effective mass spectrum.
The mass spectrum is very similar to that observed previously in the \pic
channel~\cite{re:wa1024pi},
namely, a clear peak at 1.28~GeV associated with the
$f_1$(1285), a peak at 1.45~GeV
and a broad enhancement around 2.0~GeV.
\par
A spin-parity analysis of the
\pim channel
has been performed using an isobar model and the method
described in ref.~\cite{re:wa914pi}.
Assuming that
only angular momenta up to 2 contribute,
the intermediate
states considered are
\begin{tabbing}
abddb \= mmmmmmmmm \= mmmmmmmmm \= mmmmmmmmm \= mmmmmmmmm  \= mmmm\kill
%\> $\sigma\sigma $, \> $\sigma(\pi\pi)_{S wave} $,\> $\sigma(\pi\pi)_{P wave}
%%$, \> $\sigma(\pi\pi)_{D wave} $, \> $\sigma \rho^{0} $, \\
\> $\sigma\sigma $,  \> $\sigma \rho^{0} $, \\
\>$\rho^{+}\rho^{-} $, \> $\rho^{\pm}(\pi^{\mp}\pi^{0})_{S wave} $, \>
$\rho^{\pm}(\pi^{\mp}\pi^{0})_{P wave} $, \> $\rho^{\pm}(\pi^{\mp}\pi^{0})_{D
wave} $, \\
\> $a_{1}(1260)\pi $, \> $a_{2}(1320)\pi $,
\> $f_{2}(1270)\sigma $, \> $f_{2}(1270)(\pi\pi)_{S wave} $, \\
\> $f_{2}(1270)\rho^0 $, \> $f_{2}(1270)f_2(1270) $  \\
\> $\pi^*(1300)\pi$  \\
% \linebreak
\end{tabbing}
Three parameterisations of the $\sigma$ have been tried, that of
Au, Morgan and Pennington~\cite{re:AMP}, that of Zou and
Bugg~\cite{re:zbugg} and the parameterisation found to fit the
$\sigma$ contribution in the \piz channel described above.
Three different parameterisations of the $\pi^*(1300)$ have been tried.
In the Crystal Barrel analysis two parameterisations of the
$\pi^*(1300)$ have been found, one with
M~=~1114~MeV, $\Gamma$~=~340~MeV~\cite{cb4pi} and a second with
M~=~1400~MeV, $\Gamma$~=~275~MeV~\cite{thoma}.
The third parameterisation uses the PDG values~\cite{PDG98}.
%Events corresponding to the $\eta \pi^0$ and $\omega \pi^0$
% final states have been removed by a cut on the $\pi^+\pi^-\pi^0$
% mass spectrum.
\par
Different combinations of waves and isobars have been tried and
insignificant contributions have been removed from the final fit.
The best fit is shown in
fig.~\ref{fi:3} and consists of the following waves:
$J^{PC}=1^{++}$~$\rho\rho$ with $|J_Z|$~=~1 fig.~\ref{fi:3}b),
$J^{PC}=0^{++}$~$\rho\rho $ fig.~\ref{fi:3}c),
$J^{PC}=0^{++}$~$\sigma\sigma $ fig.~\ref{fi:3}d),
$J^{PC}=2^{-+}$~$a_2(1320)\pi$  with $|J_Z|$~=~1 fig.~\ref{fi:3}e) and
$J^{PC}=2^{++}$~$f_2(1270) \pi \pi $ with $J_Z$~=~0 fig.~\ref{fi:3}f).
As in the case of the analysis of the \etapipi channel~\cite{etapipi} there
is no need for any
$J^{PC}$~=~$2^{++}$ $a_2(1320)\pi$ wave.
The $J^{PC}=0^{++}$~$\sigma\sigma $ wave is only required if the
parameterisation of the $\sigma$ found from the \piz analysis is used.
If the parameterisations of the $\sigma$
used to fit the \pic channel in our previous
publication~\cite{re:wa1024pi} are used here, the same conclusion
would be drawn, i.e. no $J^{PC}=0^{++}$~$\sigma\sigma $  wave
is required. Hence as was stated in the introduction the
parameterisation used to describe the $\sigma$ is crucial.
There is no need for any $J^{PC}$~=~$0^{++}$ $\pi^*(1300)\pi$ wave irrespective
of the parameterisation used.
\par
Superimposed on the
$J^{P}=1^{+}$~$\rho\rho $ wave shown in fig.~\ref{fi:3}b) is
a Breit Wigner convoluted with a Gaussian used to describe the
$f_1$(1285) in the fit to the \pic mass spectrum~\cite{re:wa1024pi}.
As can be seen the $f_1$(1285) is well described.
\par
The $J^{P} =0^{+} \rho\rho $ distribution
in fig.~\ref{fi:3}c) shows a peak at 1.45~GeV
together with a broad enhancement around 2~GeV.
A fit has been performed to the
$J^{P} =0^{+} \rho\rho $ amplitude in fig.~\ref{fi:3}c)
using a single channel K matrix formalism~\cite{KMATRIX}
including poles to describe the interference between the
$f_0(1370)$, the $f_0(1500)$ and a possible state at 2~GeV.
No account has been made for the $\rho \rho$ threshold in this fit.
The result of the fit is superimposed on the
$J^{P} =0^{+} \rho\rho $ distribution shown in fig.~\ref{fi:3}c) and well
describes the data. The resulting T-matrix
sheet II pole positions~\cite{sheet} for the resonances are
\begin{tabbing}
00000aaaa\=adfsfsf99ba \=Mas  == 1224 pm0 \=i12\=2400 \=pi \=1200000  \=MeV
\kill
\>$f_0(1370)$ \>M $ \; = \;$(1309$\; \pm\; $24)\>$-i$\>(163\>$\pm$\>26)\>MeV\\
\>$f_0(1500)$ \>M $ \; = \;$(1513$\; \pm\; $12)\>$-i$\>($ \;
\;58$\>$\pm$\>12)\>MeV\\
\>$f_0(2000)$ \>M $ \; = \;$(1989$\; \pm\; $22)\>$-i$\>($ \;
224$\>$\pm$\>42)\>MeV
\end{tabbing}
These parameters are consistent with the PDG~\cite{PDG98} values for the
$f_0(1370)$ and $f_0(1500)$.
\par
The $J^{P} =0^{+} \sigma\sigma $ distribution
in fig.~\ref{fi:3}d) shows a peak at 1.5~GeV.
Superimposed on fig.~\ref{fi:3}d) is what would be expected
from the $f_0(1500)$; as can be seen there is little
evidence for any $f_0(1370)$ contribution.
A result similar to what was found in the analysis of the \piz final state.
Correcting for the unseen $\rho \rho$ and $\sigma \sigma$ decay modes
the branching ratio
of $f_0(1500)$
to $\rho \rho$/$\sigma \sigma$~=~3.3~$\pm$~0.5.
\par
These results disagree with the values found by the Crystal Barrel
collaboration~\cite{cb4pi,thoma} which found a considerably larger
contribution of
$\sigma \sigma$ in the decays of the $f_0(1370)$ and $f_0(1500)$.
In order to determine the maximum $\sigma \sigma$ contribution
the parameters of the $\sigma$ would need to be changed such that
it has a similar
shape to the $\rho$. In this case
the difference between a $\rho \rho$ and
$\sigma \sigma$ decay mode would be determined by the relative
angular momentum between the pions from the decay of the $\rho$ or
$\sigma$.
Using such a parameterisation for the $\sigma$ we can determine a maximum
contribution in
the decay of the $f_0(1370)$ to $\sigma \sigma$ of 30~\%
of the $\rho \rho$ decay.
For the $f_0(1500)$ such a parameterisation of the $\sigma$ would yield
equal decay rates to $\sigma \sigma$ and $\rho \rho$.
\par
As in the case of the \etapipi channel~\cite{etapipi}, the
$J^{PC}$~=~$2^{-+}$ $a_2(1320)\pi$ wave
is consistent with being due to two resonances,
the $\eta_2(1645)$ and the $\eta_2(1870)$.
Superimposed on
fig.~\ref{fi:3}e) is the result of a fit
using two resonances to describe the $\eta_2(1645)$
and $\eta_2(1870)$.
The masses and widths determined for each resonance
are given in table~\ref{ta:a} and
are consistent with those found previously in the $\eta \pi \pi$ final
state~\cite{etapipi}.
\par
Fig.~\ref{fi:3}f)
shows the $J^{P} =2^{+}$ $f_{2}(1270)(\pi\pi)_{S wave} $ distribution
which has been fitted with
a single Breit-Wigner, with
M = 1980 $\pm$ 22 MeV and $\Gamma$ = 520 $\pm$ 50 MeV.
These values are compatible
with those coming from the fit to $f_2(1950)$ from the
\pic mass spectrum~\cite{re:wa1024pi}.
\par
Finally,
a reanalysis of the \pic channel has now been performed
using the new parameterisation of the
$\sigma$ described above. A new version of the
Monte Carlo program has also been used which has
been improved since the previous publication~\cite{re:wa1024pi}.
The \pic effective mass spectrum is shown in
fig.~\ref{fi:4}a) and consists of
1~167~089 centrally produced events.
\par
The analysis has been performed independently from the \pim
analysis and
different combinations of waves and isobars have been tried and
insignificant contributions have been removed from the final fit.
As in the original analysis~\cite{re:wa1024pi}
of the \pic system the addition of the
$J^{PC}$~=~$2^{++}$ $a_2(1320)\pi$ wave improved the
Log Likelihood in the 1.8 to 1.9 GeV mass interval
and has the effect of splitting
the $f_2(1950)$ signal. If we use the information from the
\etapipi~\cite{etapipi} and \pim analysis and reject this wave then
the best fit is shown in
fig.~\ref{fi:4} and consists of the following waves:
$J^{PC}=1^{++}$~$\rho\rho $ with $|J_Z|$~=~1 fig.~\ref{fi:4}b),
$J^{PC}=0^{++}$~$\rho\rho $ fig.~\ref{fi:4}c),
$J^{PC}=0^{++}$~$\sigma\sigma $ fig.~\ref{fi:4}d),
$J^{PC}=2^{-+}$~$a_2(1320)\pi $ with $|J_Z|$~=~1 fig.~\ref{fi:4}e) and
$J^{PC}=2^{++}$~$f_2(1270) \pi \pi $ with $J_Z$~=~0 fig.~\ref{fi:4}f).
The results are then very similar to those found for the
\pim channel.
\par
A fit to the
$J^{P} =2^{+}$ $f_{2}(1270)(\pi\pi)_{S wave} $ distribution
shown in fig.~\ref{fi:4}f) gives
M = 1940 $\pm$ 22 MeV and $\Gamma$ = 485 $\pm$ 55 MeV.
In order to get further confidence that the $f_2(1950)$
has a dominant $f_2(1270) \pi \pi $ decay mode a measurement
of its rate to \pic, \pim and \piz has been made.
The predicted ratio for an I~=~0 state would be \pic:\pim:\piz = 4~:~4~:~1.
The measured rate is 4~:~$3.7\pm0.2$~:~0.95$\pm$0.1 consistent
with the prediction. However, if the $a_2(1320)\pi$ wave
was introduced in the \pic analysis the measured \pic rate
would drop by a factor of 2.
\par
Therefore, although the Log Likelihood would indicate
the presence of a
$J^{PC}$~=~$2^{++}$ $a_2(1320)\pi$ wave in the fit
of the \pic channel,
the lack of it in the \etapipi and \pim together
with the fact that the measured decay rates would be wrong
indicates how dangerous it can be to rely on Log Likelihood
from a single channel alone to determine the existence of a decay mode.
\par
Superimposed on the
$J^{P}=1^{+}$~$\rho\rho $ wave shown in fig.~\ref{fi:4}b) is
a Breit Wigner convoluted with a Gaussian used to describe the
$f_1$(1285) in the fit to the mass spectrum.
As can be seen the $f_1$(1285) is well described.
\par
A fit to the
$J^{P} =0^{+} \rho\rho $ distribution shown
in fig.~\ref{fi:4}c) has been performed using the parametrisation used
in the \pim channel.
The T-matrix
sheet II pole positions for the resonances are
\begin{tabbing}
00000aaaa\=adfsfsf99ba \=Mas  == 1224 pm0 \=i12\=2400 \=pi \=1200000  \=MeV
\kill
\>$f_0(1370)$ \>M $ \; = \;$(1295$\; \pm\; $24)\>$-i$\>(169\>$\pm$\>26)\>MeV\\
\>$f_0(1500)$ \>M $ \; = \;$(1509$\; \pm\; $12)\>$-i$\>($ \;
\;44$\>$\pm$\>12)\>MeV\\
\>$f_0(2000)$ \>M $ \; = \;$(1995$\; \pm\; $22)\>$-i$\>($ \;
218$\>$\pm$\>42)\>MeV
\end{tabbing}
These parameters are in good agreement with those found in the
\pim final state.
\par
The $J^{P} =0^{+} \sigma\sigma $ distribution
in fig.~\ref{fi:4}d) shows a peak at 1.5~GeV.
Superimposed on fig.~\ref{fi:4}d) is what would be expected
from the $f_0(1500)$.
Similar to what was found in the \piz and \pim final states
there is little
evidence for any $f_0(1370)$ contribution.
Correcting for the unseen $\rho \rho$ and $\sigma \sigma$ decay modes
the branching ratio
of $f_0(1500)$
to $\rho \rho$/$\sigma \sigma$~=~2.6~$\pm$~0.4 which is
consistent with the value found from the \pim final state.
As in the \pim channel
a parameterisation of the $\sigma$ has been used with a shape
similar to that of the $\rho(770)$ which
give an upper limit to
the $\sigma \sigma$ contribution
in the decays of the $f_0(1370)$ and $f_0(1500)$.
For the $f_0(1370)$ the maximum contribution is 25~\%
of the $\rho \rho$ decay.
For the $f_0(1500)$
the maximum $\sigma \sigma$ contribution
is 80~\%
of the $\rho \rho$ decay.
\par
Fig.~\ref{fi:4}e) shows the
$J^{PC}$~=~$2^{-+}$ $a_2(1320)\pi$ wave.
Superimposed
is the result of a fit
using two resonances to describe the $\eta_2(1645)$
and $\eta_2(1870)$.
The masses and widths determined for each resonance
are given in table~\ref{ta:a} and as can be seen
are consistent with those found for the $\eta \pi \pi$~\cite{etapipi} and
\pim final states.
Table~\ref{ta:a} also gives the combined masses and widths
from all the decay modes, including those found in the
\etapipi channel~\cite{etapipi}, where the common systematic errors
have been taken into account.
% The $\eta_2(1645)$ is found to
% have a mass of 1620~$\pm$~12~MeV
% and a width of 170~$\pm$~22~MeV and
% the $\eta_2(1870)$ to have
% a mass
% of 1840~$\pm$~23~MeV
% and a width of 220~$\pm$~32~MeV.
In both the \pim and \pic channels there is also evidence
for a
$J^{P}=2^{+}$~$\rho\rho $ wave.
Except for evidence for the $f_2(1270)$ this wave is a broad,
structure-less distribution.
\par
In order to gain further confidence in the results, we have measured
the
rate to \pic, \pim and \piz for the other waves
observed.
As can be seen from table~\ref{ta:b} the results
obtained are consistent with the expectations.
\par
In the analyses presented in this paper,
the decay modes of the $f_0(1370)$ and $f_0(1500)$ are in disagreement
with what has been found by the Crystal Barrel collaboration.
In order to demonstrate the clear need for a sizable $\rho \rho$
contribution to the peak at 1.45 GeV
a study has been made of the
2$\pi$ and 3$\pi$ effective mass spectra  for
the mass range 1.4~$\leq$~M(4$\pi$)~$\leq$~1.5 GeV
and for $dP_T$~$\leq$~0.2~GeV for which the peak at
1.45 GeV is most prominent~\cite{re:wa1024pi}.
Fig.~\ref{fi:5}a) and b) show the
$\pi^+\pi^-$ and $\pi^+\pi^-\pi^\pm$ mass spectra respectively
from this region for the \pic final state. Superimposed on the
mass spectra as a dashed line is what would be expected from
a $\rho \rho$ final state and shaded is the contribution from a
$\sigma \sigma$ final state using the parametrisation of the $\sigma$
found from the \piz channel.
The solid curve shows the sum of the two
contributions and as can be seen, apart from an excess of events in the
$K^0$ mass region of the $2\pi$ mass spectrum,
the simulation well describes the experimental $\pi^+\pi^-$ and
$\pi^+\pi^-\pi^\pm$ mass spectra. Therefore it can be seen that a
strong $\rho \rho$ contribution is needed in this mass region
and there is no need for any $\pi^*(1300)$ contribution to the
$3\pi$ mass spectrum. A similar conclusion is found from an analysis for the
\pim final state.
\par
In summary,
there is evidence for an $a_2(1320) \pi$ decay mode of the
$\eta_2(1645)$ and $\eta_2(1870)$ in the \pim and \pic
final states.
The $f_2(1950)$ is consistent with being a single resonance with
a dominant $f_2(1270) \pi \pi $ decay mode.
The $f_0(1370)$ is found to decay dominantly to $\rho \rho$ while
the $f_0(1500)$ is found to decay to $\rho \rho$ and $\sigma \sigma$.
\begin{center}
{\bf Acknowledgements}
\end{center}
\par
This work is supported, in part, by grants from
the British Particle Physics and Astronomy Research Council,
the British Royal Society,
the Ministry of Education, Science, Sports and Culture of Japan
(grants no. 07044098 and 1004100), the French Programme International
de Cooperation Scientifique (grant no. 576)
and
the Russian Foundation for Basic Research
(grants 96-15-96633 and 98-02-22032).

\newpage
{ \large \bf Tables \rm}
\begin{table}[h]
\caption{Parameters of the $\eta_2(1645)$ and $\eta_2(1870)$}
\label{ta:a}
\vspace{1in}
\begin{center}
\begin{tabular}{|c|c|c|c|} \hline
  & & & \\
 Resonance &Final state&Mass (MeV) & Width (MeV) \\
 & & &  \\
 & & &  \\ \hline
 & & &  \\
$\eta_{2}(1645)$ & \pim &1621 $\pm$ 11 &168 $\pm$ 23  \\
 &  & &  \\
$\eta_{2}(1645)$ & \pic &1624 $\pm$ 12 &173 $\pm$ 22  \\
 & & &  \\ \hline
 &  & &  \\
$\eta_{2}(1645)$ & Combined &1617 $\pm$ 8 &177 $\pm$ 18  \\
 & & &  \\ \hline
 & & &  \\
$\eta_{2}(1870)$ & \pim &1862 $\pm$ 19 &222 $\pm$ 28  \\
 &  & &  \\
$\eta_{2}(1870)$ & \pic &1843 $\pm$ 23 &226 $\pm$ 32  \\
 & & &  \\ \hline
 &  & &  \\
$\eta_{2}(1870)$ & Combined &1844 $\pm$ 13 &228 $\pm$ 23  \\
 & & &  \\ \hline
\end{tabular}
\end{center}
\end{table}
% \newpage
\begin{table}[h]
\caption{The predicted rate for I~=~0 and measured rate to
\pic : \pim : \piz for different waves.}
\label{ta:b}
\vspace{1in}
\begin{center}
\begin{tabular}{|c|c|c|} \hline
  & & \\
 Wave &Predicted ratio& Measured ratio \\
 & &  \\
 & &  \\ \hline
 & &  \\
$2^{++}$ $f_2(1270)\pi \pi $ & 4 : 4 : 1 & 4 : 3.7$\pm$0.2 : 0.95$\pm$0.2  \\
  & &  \\
$1^{++}$ $ \rho \rho$ & 1 : 2 : 0 & 1 : 2.1$\pm$0.2 : 0.  \\
  & &  \\
$0^{++}$ $ \rho \rho$ & 1 : 2 : 0 & 1 : 1.9$\pm$0.2 : 0.  \\
  & &  \\
$2^{-+}$ $ a_2(1320)\pi$ & 1 : 2 : 0 & 1 : 2.0$\pm$0.2 : 0.  \\
  & &  \\
$0^{++}$ $ \sigma \sigma$ & 4 : 4 : 1 & 4 : 3.9$\pm$0.2 : 1.2$\pm$0.3 \\
 & &  \\ \hline
\end{tabular}
\end{center}
\end{table}
\clearpage
{ \large \bf Figures \rm}
\begin{figure}[h]
\caption{The \piz channel.
a) M($\gamma \gamma$) when the other three $\gamma \gamma$ pairs lie
in the $\pi^0$ region,
b) M(\piz) and c) M($\pi^0\pi^0$) for 1.3~$\le$~M(\piz)~$\le$~1.5~GeV.
Results of the spin analysis
d) the $J^{PC}$~=~$0^{++}$ $\sigma \sigma$ wave and
e) the $J^{PC}$~=~$2^{++}$ $f_2(1270) \pi \pi $ wave with fits
described in the text.
}
\label{fi:1}
\end{figure}
\begin{figure}[h]
\caption{The \pim channel.
a) The total mass spectrum,
b)~$1^{++}$~$\rho \rho$,
c)~$0^{++}$~$\rho \rho$,
d)~$0^{++}$~$\sigma  \sigma$,
e)~$2^{-+}$~$a_{2}(1320)\pi$ and
f)~$2^{++}$~$f_{2}(1270)\pi\pi$.
The superimposed curves are the resonance contributions coming from
the fits described in the text.}
\label{fi:3}
\end{figure}
\begin{figure}[h]
\caption{The \pic channel.
a) The total mass spectrum,
b)~$1^{++}$~$\rho \rho$,
c)~$0^{++}$~$\rho \rho$,
d)~$0^{++}$~$\sigma  \sigma$,
e)~$2^{-+}$~$a_{2}(1320)\pi$ and
f)~$2^{++}$~$f_{2}(1270)\pi\pi$.
The superimposed curves are the resonance contributions coming from
the fits described in the text.}
\label{fi:4}
\end{figure}
\begin{figure}[h]
\caption{
a) The $\pi^+\pi^-$ and
b) $\pi^+\pi^-\pi^\pm$ mass spectra for the \pic
final state in the mass range 1.4~$\leq$~M(\pic)~$\leq$~1.5 GeV
and with $dP_T$~$\leq$~0.2~GeV. The superimposed histograms
are for a $\rho \rho$ (dashed) and $\sigma \sigma$ (shaded)
final state. The solid line represents the sum of the
$\rho \rho$ (90 \%) and $\sigma \sigma$ (10 \%) contributions.}
\label{fi:5}
\end{figure}
\begin{center}
\epsfig{figure=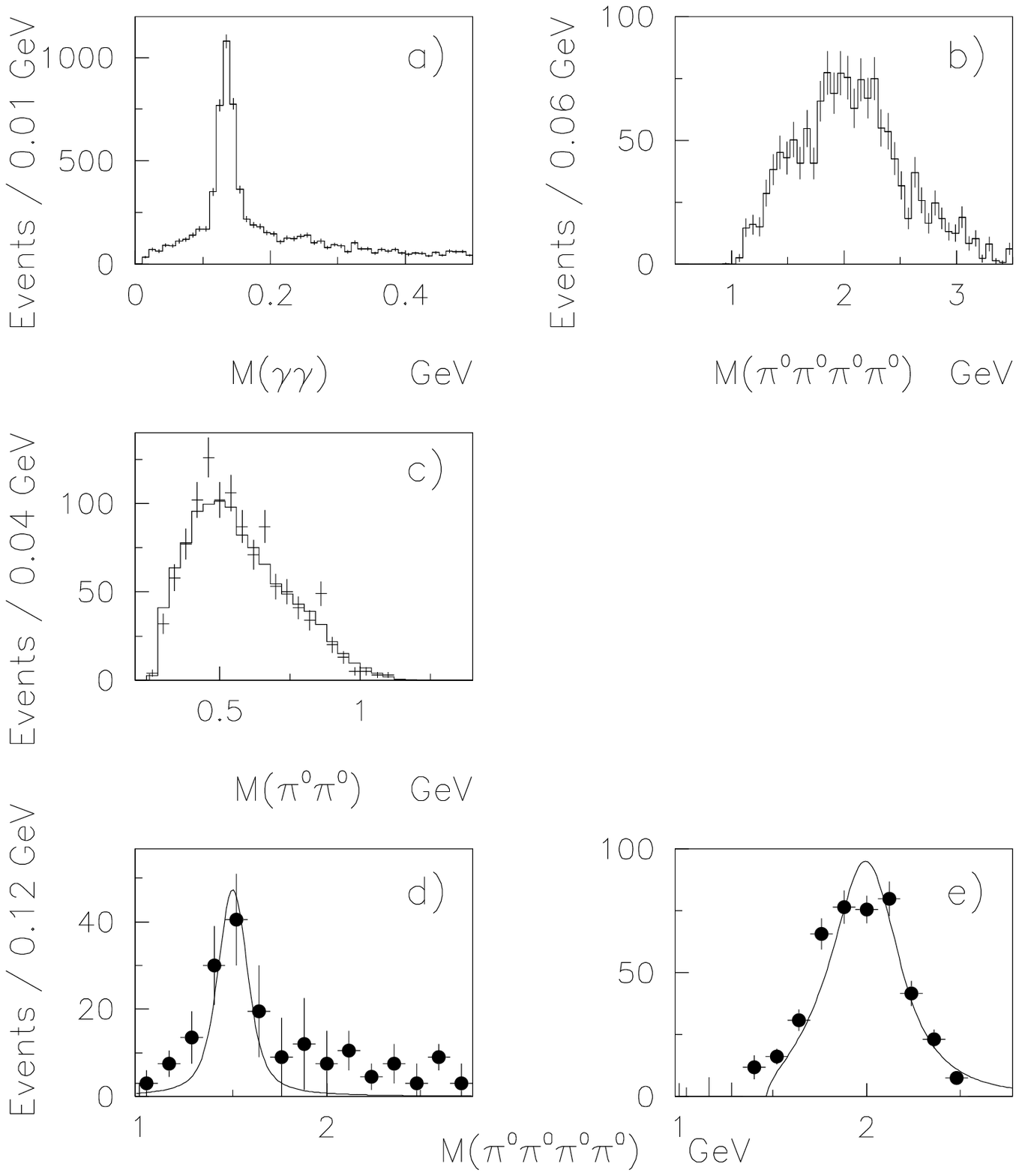,height=22cm,width=17cm}
\end{center}
\begin{center} {Figure 1} \end{center}
\newpage
\begin{center}
\epsfig{figure=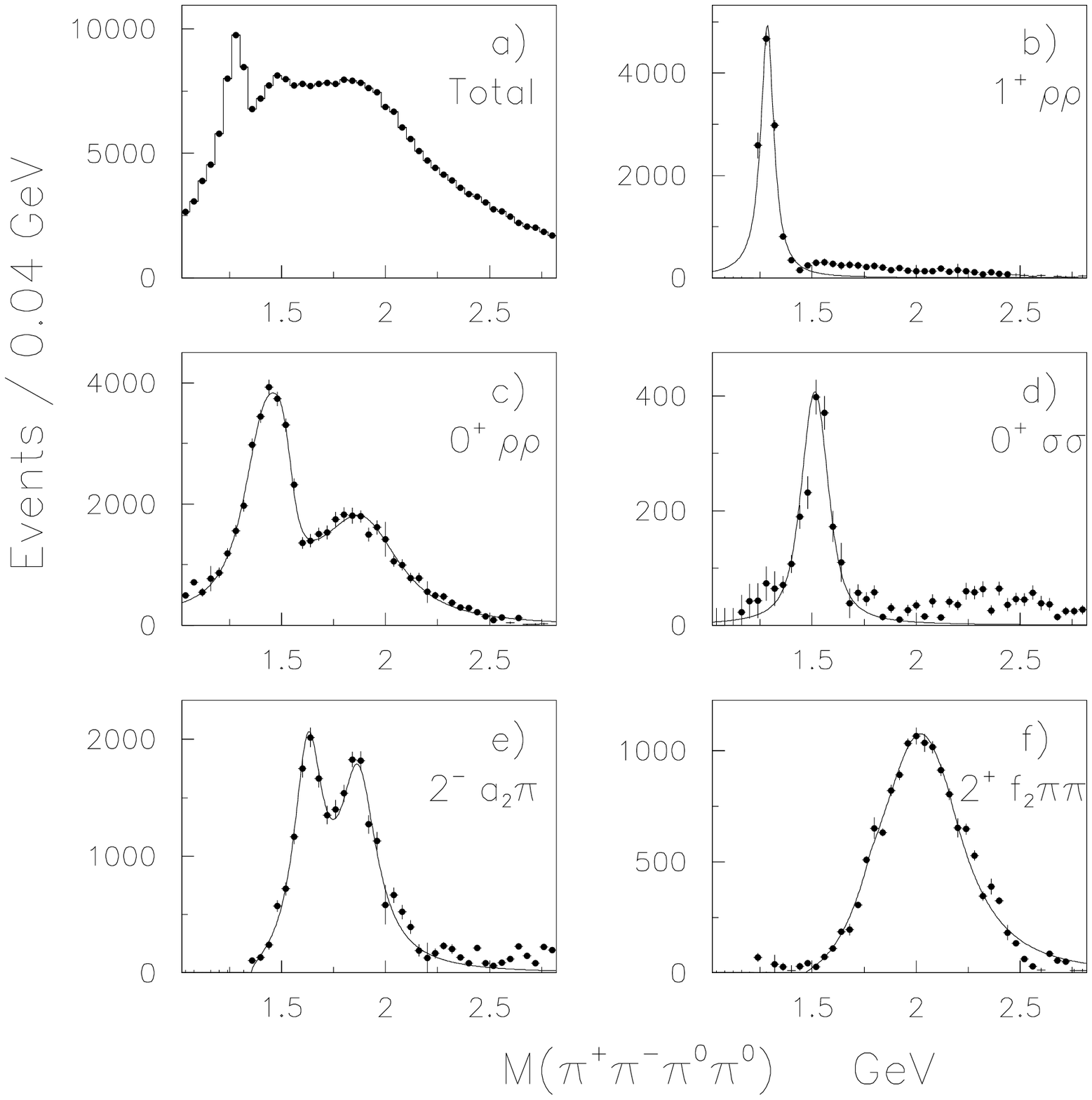,height=22cm,width=17cm}
\end{center}
\begin{center} {Figure 2} \end{center}
\newpage
\begin{center}
\epsfig{figure=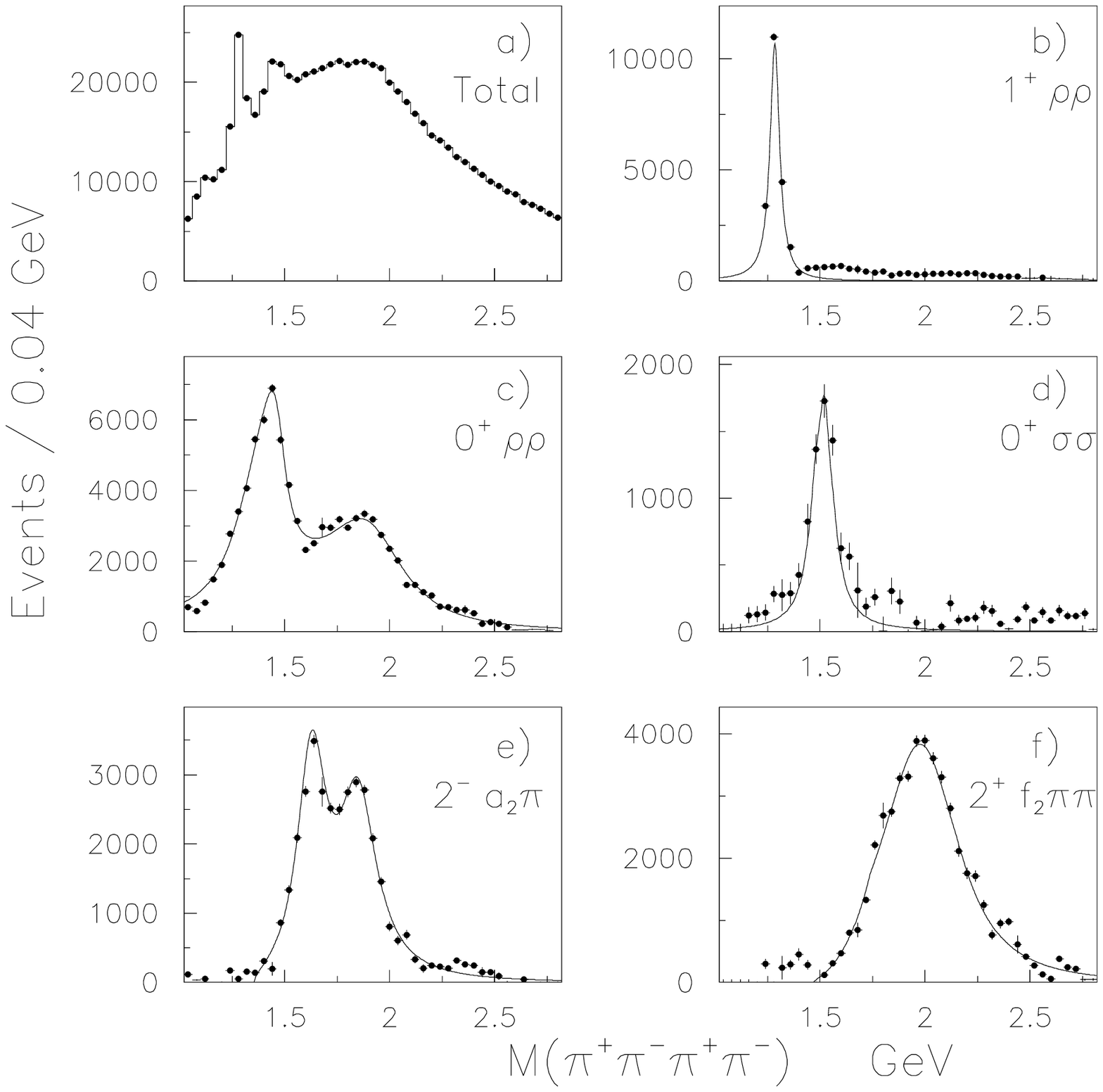,height=22cm,width=17cm}
\end{center}
\begin{center} {Figure 3} \end{center}
\newpage
\begin{center}
\epsfig{figure=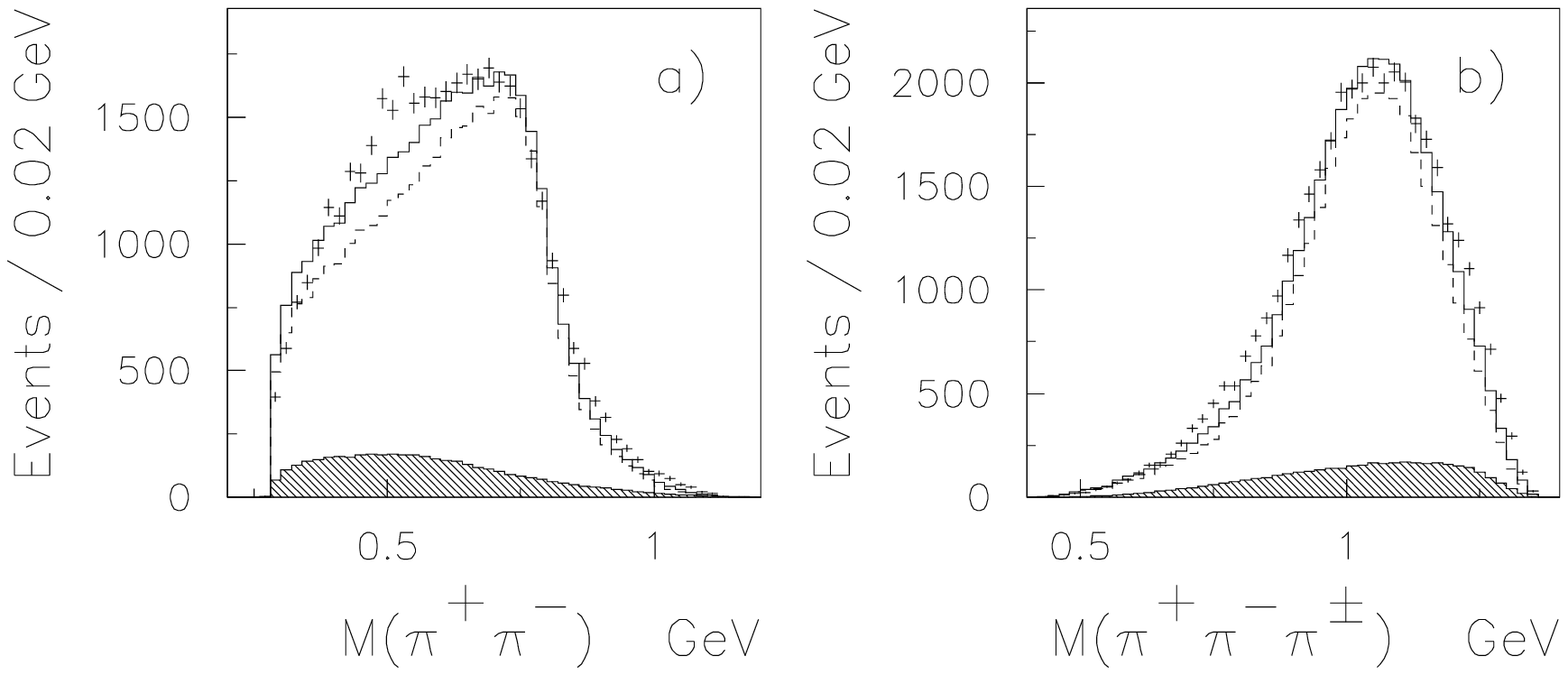,height=22cm,width=17cm}
\end{center}
\begin{center} {Figure 4} \end{center}
\end{document}